\input{aipcheck}

\documentclass[
    ,final            
  ]
  {aipproc}

\def\beq {\begin{equation}}
\def\eeq {\end{equation}}
\def\bea {\begin{eqnarray}}
\def\eea {\end{eqnarray}}
\def\ni {\noindent}
\def\nn {\nonumber}

\def\ul {\underline}

\def\lb {\left[ }
\def\rb {\right] }

\def\rar {\rightarrow}

\def\kb {\bar{\k}}

\def\Ob {\bar{\Omega}}
\def\Rb {\bar{R}}

\def\sp {\!+\!}
\def\sm {\!-\!}
\def\cd {\!\cdot\!}

\def\so {{_{1/2}}}
\def\st {{_{3/2}}}

\def\cA {{\cal{A}}}
\def\cK {{\cal{K}}}

\def\d {\delta}

\def\k {\kappa}

\def\m {\mu}

\def\o {\omega}
\def\p {\pi}
\def\s {\sigma}

\layoutstyle{6x9}

\begin{document}

\title{$D^+ \!\rightarrow K^- \, \p^+ \, \p^+ \;$:
the low-energy sector}


\author{D. R. Boito*, P. C. Magalh\~aes, 
\ul{M. R. Robilotta} and G. R. S. Zarnauskas}
{  address=
{Instituto de F\'{i}sica, Universidade de S\~ao Paulo,
S\~ao Paulo, SP, Brazil\\
{*}Grup de F\'isica Te\'orica and IFAE, Universitat Aut\`onoma de Barcelona,
E-08193 Bellaterra (Barcelona), Spain. }}

\begin{abstract}
An effective $SU(3)\times SU(3)$ chiral lagrangian, which includes scalar
resonances, is used to describe the process $D^+ \! \rar K^- \p^+ \p^+$
at low-energies.
Our main result is a set of five $S$-wave amplitudes, suited to be used 
in analyses of production data.
\end{abstract}

\maketitle


\section{introduction}

The pioneering E791 experiment\cite{E791} has shown that heavy-meson 
decays are reliable sources of information about scalar resonances. 
Data about these decays are encoded into Dalitz plots and have 
to be interpreted with the help of theoretical ans\"atze which,
normally, rely on Breit-Wigner expressions.
In the framework of field theory, one knows that expressions of this kind
arise naturally when two-body interaction kernels are unitarized.
However, usual Breit-Wigner functions are problematic, since they are based 
on kernels which do not describe well the meson-meson amplitude close 
to threshold.
We propose to cure this problem by means of generalized Breit-Wigner 
expressions,  based on interaction kernels which
comply with low-energy chiral theorems.
These alternative trial functions can be used as tools in 
analyses of the decay $D^+ \! \rar K^- \p^+ \p^+ $.

Our theoretical description is based on an effective $SU(3)\times SU(3)$ 
chiral lagrangian, which includes scalar resonances and
also allows a consistent treatment of the primary weak vertex.
Final expressions represent a compromise between reliability and 
simplicity, so that they could be employed directly in data analyses.

\section{results}

The decay $D^+(P) \rar \p^+(q) \p^+(q') K^-(k)$ involves both the weak 
quark transition $c \rar s \, W^+$ and strong processes associated 
with final state interactions.
We only consider hadronic degrees of freedom and need the following
weak vertices:
$\; (D \rar \p \, K \, W)_a \;,$ 
$\; (D\rar \kb \, W)_a \;,$
$\; (W \rar \p)_a \;,$
$\; (D \rar K \, W)_v \;,$
and $\; (W \rar \p \, \p)_v \;,$
where $\kb$ is the {\em kappa}-resonance and the labels $v$ and $a$ 
refer to vector and axial currents involved in $W$-couplings.
Strong interactions determine most structures 
observed in Dalitz plots and, in particular, the widths of resonances.
The final state one is considering contains three mesons and 
a full treatment of their interactions is impossible.
Therefore we work in the approximation in which one of the final mesons 
acts as a spectator.
As the emerging pions have isospin 2, their interactions can be neglected.
Strong processes are then restricted to the $\p K$ subsystem 
and described using the leading chiral $SU(3)\times SU(3)$ 
lagrangian given by Gasser and Leutwyler\cite{GL}, 
complemented with resonance couplings 
from the work of Ecker, Gasser, Pich and De Rafael\cite{EGPR}.

\begin{figure}[h] 
\includegraphics[width=0.8\columnwidth,angle=0]{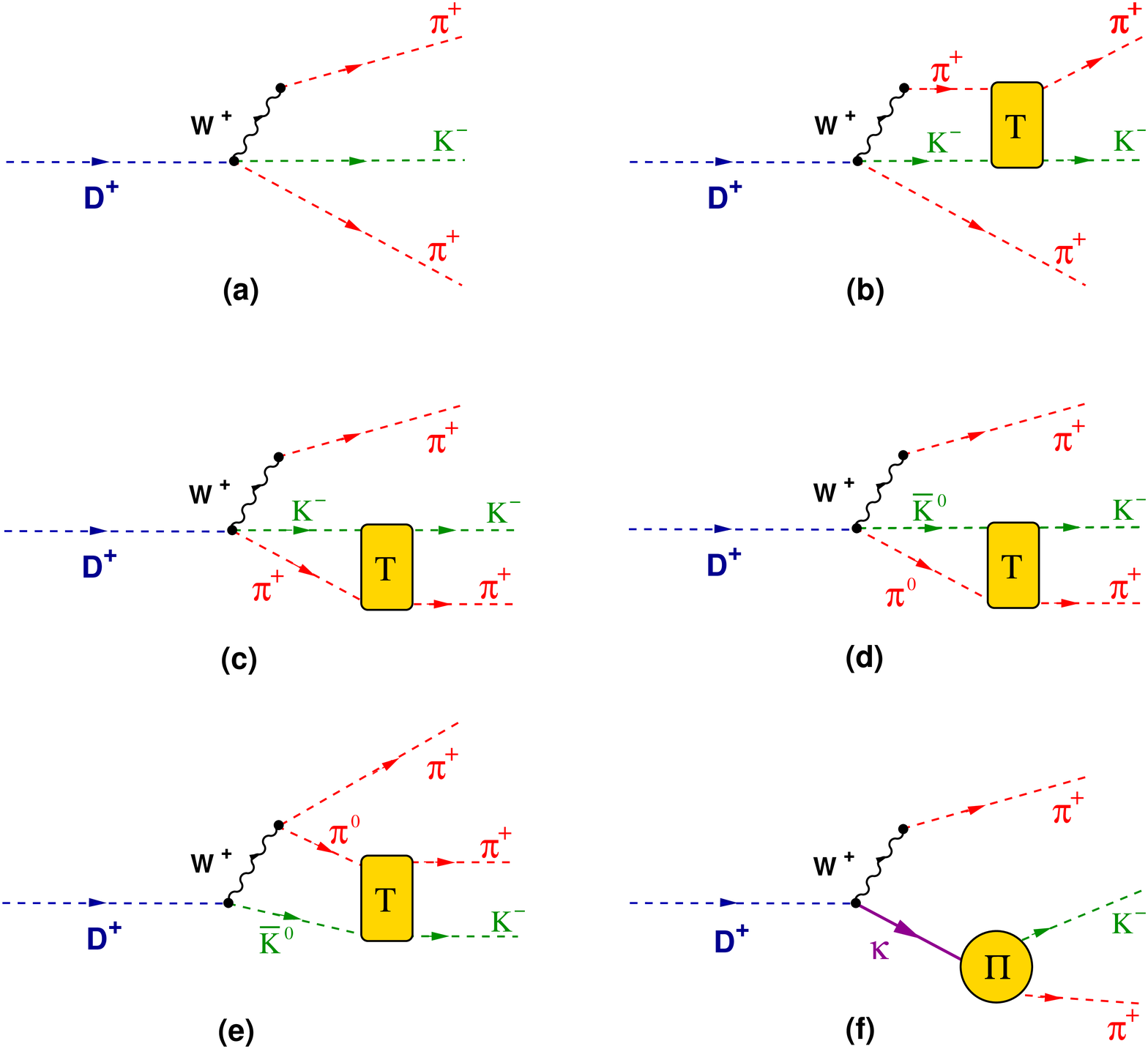}
\caption{(Color online) Diagrams contributing to the decay
$D^+ \rar K^- \p^+ \p^+ $; (a) corresponds to a direct process,
(b-e) involve the $\p K$ scattering amplitude $T$,
and (f) depends on the production amplitude $\Pi$.}
\label{F2}
\end{figure}

\begin{figure}[h]
\includegraphics[width=0.7\columnwidth,angle=0]{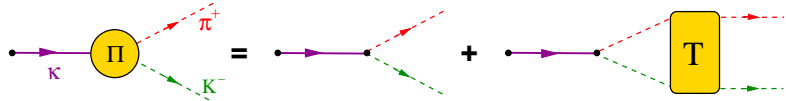}
\caption{ (Color online) Resonance  propagation (full line) and decay 
into $\p^+ K^-$ 
(dashed lines); $T$ is the unitary $I=1/2$ scattering amplitude.}
\label{F11}
\end{figure}

Our model is defined by fig.\ref{F2}.
Diagram (a) represents a non-resonant background, since the outgoing 
mesons reach the detectors without interacting.
The other diagrams involve the strong amplitudes $T$ and $\Pi$.
The former describes elastic scattering and is unitary.
The latter implements the propagation and decay of the $\k$ produced 
directly at the weak vertex\cite{BR07} and has the dynamical structure
shown in fig.\ref{F11}.

Individual contributions from diagrams of fig.\ref{F2}
to the $D^+$ decay width read 
\bea
\cA_a(\mu_{K \p}^2, \mu_{K\p'}^2) \!\!\!&=&\!\!\! 
\frac{1}{6\sqrt{2}}\; [\d_a] \; G_F\cos^2\theta_C \;
[(M_D^2 \sp 2M_\p^2 \sp M_K^2 \sm \m_{K\p}^2)
\nn\\
&+&\!\!\! (M_D^2 \sp 2M_\p^2 \sp M_K^2 \sm \m_{K\p'}^2)]\;,
\label{L.1}\\[2mm]
\cA_b(\mu_{K\p'}^2) \!\!\!&=&\!\!\! 
-\,\frac{1}{9\sqrt{2}}\; [\d_a] \; G_F\cos^2\theta_C \;
\lb (P\cd q \sp M_\p^2) -\frac{(P\cd q \sm M_\p^2) \, (M_K^2 \sm M_\p^2)}
{M_D^2 \sp M_\p^2 \sm 2\,P \cd q} \rb 
\nn\\[2mm]
&\times& \!\!\! 
\lb 2\; \Ob_\so \; T_\so(\mu_{K \p'}^2) 
+ \Ob_\st \; T_\st(\mu_{K \p'}^2) \rb \;,
\label{L.2}\\[2mm]
\cA_{c + d}(\mu_{K\p}^2) \!\!\!&=&\!\!\! 
-\,\frac{1}{3\sqrt{2}}\; [\d_a] \; G_F\cos^2\theta_C \;
\lb (P\cd q' \sp M_\p^2) -\frac{(P\cd q' \sm M_\p^2) \, (M_K^2 \sm M_\p^2)}
{M_D^2 \sp M_\p^2 \sm 2\,P \cd q'} \rb 
\nn\\
&\times& \!\!\! 
\Ob_\so \; T_\so(\mu_{K\p}^2) \;,
\label{L.3}\\[2mm]
\cA_e(\mu_{K\p}^2) \!\!\!&=&\!\!\!  -\,\frac{\sqrt{2}}{3}\; 
[\d_b] \; G_F\cos^2\theta_C \;
\lb (M_D^2 \sm 3\, P\cd q') 
-\frac{(M_D^2 \sm P\cd q') \, (M_K^2 \sm M_\p^2)}
{M_D^2 \sp M_\p^2 \sm 2\,P \cd q'} \rb 
\nn\\[2mm]
&\times& \!\!\! 
\lb \Ob_\so \; T_\so(\mu_{K\p}^2) 
- \Ob_\st \; T_\st(\mu_{K\p}^2) \rb \;,
\label{L.4}\\[2mm]
\cA_f(\mu_{K \p}^2) \!\!\!&=&\!\!\!  - 4 \sqrt{3} \, 
[\d_c] \; G_F\cos^2\theta_C \;  
\frac{P\cd q'}{\m_{K\p}^2 - m_\k^2} \; [c_d/F^2]
\label{L.5}\\[2mm]
&\times& \!\!\!
\lb c_d \, (\mu_{K \p}^2 \sm M_\p^2 \sm M_K^2) 
+ c_m \, (4\,M_K^2 \sp 5\,M_\p^2)/6 \rb\;
\lb 1 - \Ob_\so \, T_\so(\mu_{K\p}^2) \rb \;, 
\nn
\eea

\ni
where $G_F$ and $\theta_C$ are the Fermi coupling constant and
the Cabibbo angle, $F$ is the pseudoscalar decay constant,
$c_d$ and $c_m$ describe two possible $\k\p K$ couplings,
and the $[\d_i]$ implement $SU(4)$ symmetry breaking at the
weak vertices.   
One has two possible $\p K$ subsystems and their invariant masses are
$\m_{K\p}^2=(q+k)^2$ and $\m_{K\p'}^2=(q'+k)^2$.
In the construction of Dalitz plots, one first evaluates 
$\cA=[\cA_a + \cdots \cA_f]$ and then symmetrize with respect 
to $\m_{K\p}^2 \leftrightarrow \m_{K\p'}^2$.
Allowed values for these invariant masses lie in the interval
$0.40\,$GeV$^2 \leq \m_{K\p}^2,\m_{K\p'}^2 \leq 2.99\,$GeV$^2$.

The functions $\Ob$ and $T$ are important substructures in our results and 
describe respectively two-meson propagation and $\p K$ elastic scattering.
The labels $1/2$ and $3/2$ refer to $\p K$ isospin channels
and we recall that only the former can couple with a $s$-channel $\k$.
The main features of these functions are summarized below.

\section{scattering amplitude}

\begin{figure}[ht] 
\includegraphics[width=1.0\columnwidth,angle=0]{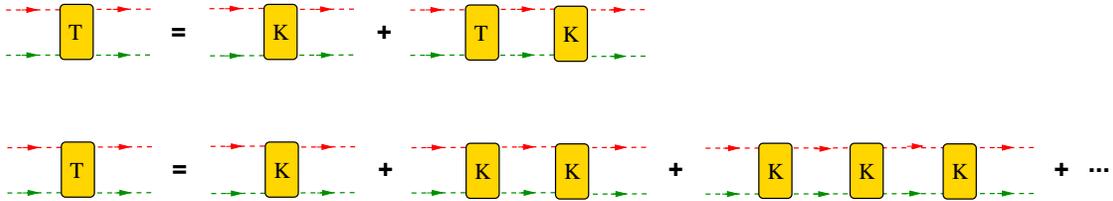}
\vspace{-8mm}
\caption{(Color online) Bethe-Salpeter equation for the elastic $\p K$
amplitude:  full equation (top) and perturbative solution (bottom).}
\label{F6}
\end{figure}
 
The elastic $\p K$ scattering amplitudes $T_I$ are derived by means 
of the Bethe-Salpeter equation, represented in figs.\ref{F6},
together with its perturbative solution.
This geometrical series involves the two-body irreducible kernels $\cK_I$
and the regularized two-meson propagators $\Ob_I$.
As pointed out by Oller and Oset\cite{OO97}, at low-energies,
the Bethe-Salpeter equation can be solved exactly,
and one has 
\beq
T_I = \frac{\cK_I}{1 + \Ob_I \, \cK_I} \;.
\label{L.6}
\eeq
Propagators are complex above threshold and given by 
$\Ob_I = \Rb_I + i\, I$.
The scattering amplitudes are unitary and can be rewritten as 
\beq
T_I = \frac{16 \p}{\rho} \,  \sin \d_I \, e^{i \d_I} 
\;\;\; \leftrightarrow \;\;\;
\tan\d_I = -\, \frac{I\;\cK_I}{1 + \Rb_I\;\cK_I}\;.
\label{L.7}
\eeq
\ni
where $\d_I$ are real phase shifts and 
$\rho=\sqrt{1 - 2\, (M_K^2 \sp M_\p^2)/s+ (M_K^2 \sm M_\p^2)^2/s^2}$.
Our unitarization of $T_I$ generalizes the $K$-matrix approximation,
which amounts to neglecting the real parts of $\Ob_I$.

\begin{figure}[h] 
\includegraphics[width=1.0\columnwidth,angle=0]{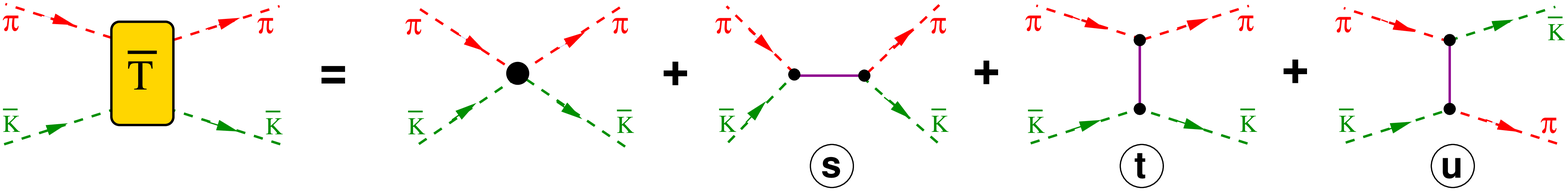}
\vspace{-12mm}
\caption{(Color online) Tree-level $\pi K$ amplitude;
dashed and full lines represent respectively pseudoscalar mesons 
and  scalar resonances.}
\label{F3}
\end{figure}

The dynamical content of the scattering amplitudes is incorporated into
the kernels $\cK_I$, which are obtained by projecting out the
$S$-wave content from the diagrams of fig.\ref{F3}.
Contributions from $t$ and $u$ channels are very small and can be neglected.
Our chiral lagrangians then yield
\bea
\cK_\so \!\! &=& \!\!  
\frac{1}{4 \,F^2} \,\lb 4\,s - 3\,\rho/2 - 4\,(M_\p^2 + M_K^2)\rb
\nn\\[2mm]
&-& \!\! \frac{3}{4}\; \frac{1}{s \sm m_\k^2}\; \frac{4}{F^4}\;
[c_d\;(s \sm M_\p^2 \sm M_K^2) + c_m \;(4\,M_K^2 \sp 5\,M_\p^2)/6 ]^2 \;,
\label{L.8}\\[2mm]
\cK_\st \!\! &=& \!\!  
-\, \frac{1}{2 \,F^2} \,\lb s - (M_\p^2 + M_K^2)\rb \;.
\label{L.9}
\eea

\section{two-meson propagator}

The propagator for a system with total momentum $X= q \sp k$, is
\beq
\Omega = i \,\int \frac{d^4 \ell}{(2 \p)^2} \;
\frac{1}
{[(\ell \sp X/2)^2 \sm M_\p^2]\;[(\ell \sm X/2)^2 \sm M_K^2]}
\label{L.10} 
\eeq
\ni
and can be written 
as $\Omega = -(1/16 \p^2) \lb L(s) \sp \Lambda_\infty \rb$,
where the function $L(s)$ is given below and $\Lambda_\infty$
is an infinite constant that has to be removed by renormalization.
Above threshold, one has 
\bea
&& L(s) = \rho \log \lb \frac{1 - \s}{1 + \s}\rb 
- 2 + \frac{(M_K^2 -M_\p^2)}{s} \,\log \frac{M_K}{M_\p}
+ i \pi \rho\;,
\nn\\[2mm]
&& \s = \sqrt{|s \sm (M_K \sp M_\p)^2|/
|s \sm (M_K \sm M_\p)^2 | } \;.
\label{L.11}
\eea
The renormalized propagators are given by 
$\Ob_I = -(1/16 \p^2) \lb L(s) \sp c_I \rb $
and the constants $c_I$ depend on the isospin channel.
They are chosen by tuning the predicted phase shifts, eq.(\ref{L.7}), 
to experimental results at a given point $s=s_I$.
For the channel $I=1/2$, the constant $c_\so$ is chosen at the point 
$s_\so = m_\k^2$, where the experimental phase is $\p/2$.
One writes
\beq
\Rb_\so (s) = -\frac{1}{16\p^2} \; \Re\! \lb L(s) - L(m_\k^2) \rb 
\label{L.12}
\eeq

\ni
and, by construction, $\Rb_\so(m_\k^2)=0$.
In the $I=3/2$ channel, fit to data requires $c_\st\sim 14.5$.
One notes that the functions $\Ob_I$ introduce other phases into the problem, 
given by $\tan \omega_I(s) \equiv I/\bar R_I$.

\section{summary}

\begin{figure}[h] 
\hspace*{-25mm}
\includegraphics[width=.6 \columnwidth,angle=0]{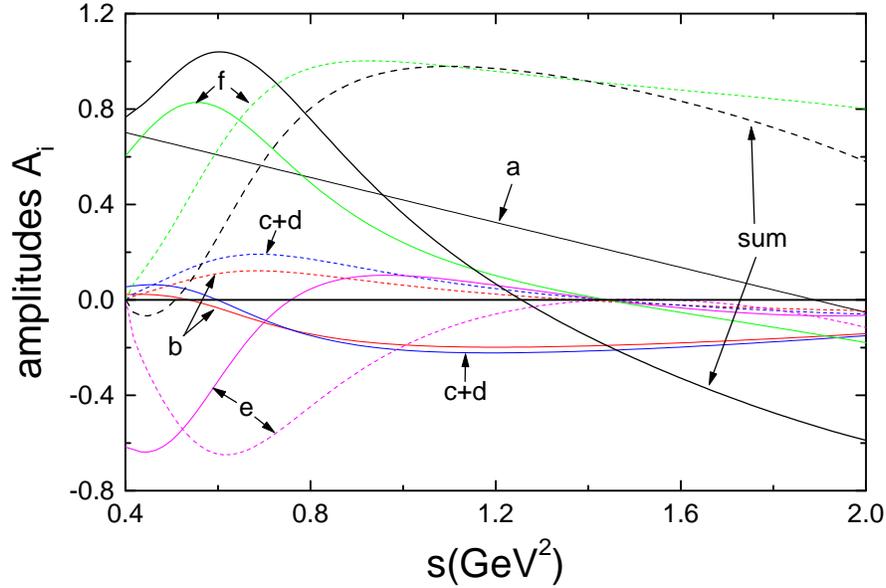}
\vspace{-110mm}
\caption{(Color online) Full (black) and partial contributions to the 
real (continuous line) and imaginary (dashed line)
components of the amplitudes $\cA_i$;
the vertical scale has to be multiplied by the weak factor $G_F\cos^2\theta_C$.}
\vspace{-50mm}
\label{F14}
\end{figure}

The purpose of this work is to map the relevant degrees of freedom of 
the amplitude $D^+ \rar K^- \p^+ \p^+$ at low energies.
This is achieved in eqs.(\ref{L.1}-\ref{L.5}), which represent
individual contributions from the diagrams shown in fig.\ref{F2}.
Masses and coupling constants in these expressions are kept free, 
so that their values can be extracted from experiment.
However, in order to discuss qualitative features of our results, 
we need to fix somehow these free parameters.
In this case, we choose:
$F=0.093\,$GeV, $m_\k=1.2\,$GeV,
$(c_d; c_m)= (3.2; 4.2)\times 10^{-2}$ GeV\cite{EGPR}
and $\d_a=\d_b=\d_c=1$.
We show, in fig.\ref{F14}, the  real and imaginary components of the 
amplitudes $\cA_i$ and note that the magnitudes of all contributions
are comparable.
Nevertheless, it is also possible to see that diagrams (a) and (f),
which describe the non-resonant background and the direct production 
of the resonance at the weak vertex, are especially important.
The latter is very sensitive to the resonance coupling constant $c_d$ 
and we take this conclusion with reserve.

In our calculation, the non-resonating background is given by 
a single and very simple diagram.
It does not contain loops and hence $\cA_a$ is necessarily a 
{\em real} function.
As the inclusion of loops is associated with higher orders in 
the chiral expansion, we  conclude that, at low-energies, 
{\em there is no phase associated with the background}. 
It is also worth noting that $\cA_a$ is linear in the invariant masses 
and its distribution over a Dalitz plot will not be uniform, 
as sometimes assumed.

Other diagrams are complex and the presence of imaginary components can
be unambiguously traced back to loops in the propagators $\Ob_I$.
By means of eq.(\ref{L.6}), they blend with the real kernels $\cK_I$
and give rise to unitary $\p K$ scattering amplitudes.
Direct inspection of eqs.(\ref{L.2}-\ref{L.4}) shows that their phases 
depend on combinations of $\d_I$ and $\o_I$.
In digram (f), representing the direct production of the $\k$-resonance
at the weak vertex, the phase is just $\d_\so$, as discussed in 
in ref.\cite{BR07}.

We assume our results for the $\cA_i$ to be reliable up to values of 
$\m_{K\p}^2$ just above the point
$s=m_\k^2$, corresponding to $\d_\so=\p/2$.
In the case of fig.\ref{F14}, this means invariant masses 
between $0.4\,$GeV$^2$ and $1.6\,$GeV$^2$.
Constraints imposed by chiral symmetry are very important
at the lower end of this interval.
Full details of these calculations will be presented elsewhere.

\begin{theacknowledgments}
It is our pleasure to thank George Rupp and collaborators for the very nice 
meeting, for the friendly hospitality and for the kindness with our group;
P.C.M. and G.R.S.Z. also thank the support for local expenses.
This work was supported by FAPESP (Brazilian Agency);
the work by DRB  is supported in part by a FPI scholarship of the
Ministerio de Educaci\'on y Ciencia under grant FPA2005-02211,
the EU Contract No.~MRTN-CT-2006-035482, ``FLAVIAnet'',
and the Spanish Consolider-Ingenio 2010 Programme CPAN
(CSD2007-00042).

\end{theacknowledgments}


\end{document}